\newcommand{\myname}{Luis Guillermo Natera Orozco}
\newcommand{\myemail}{Corresponding author: misz@itu.dk}
\newcommand{\myaffiliation}{Department of Network and Data Science\\Central European University, H-1051 Budapest, Hungary}
\newcommand{\paperdate}{\today}
\newcommand{\papertitle}{Data-driven strategies for optimal bicycle network growth}
\newcommand{\paperkeywords}{multiplex networks, sustainable transport, bicycle infrastructure, network growth}

\RequirePackage[l2tabu,orthodox]{nag}   % warn if using any obsolete or outdated commands
\documentclass[11pt,onecolumn]{article} % document style

\usepackage[T1]{fontenc}                % output T1 font encoding (8-bit) so accented characters are a single glyph
\usepackage[utf8]{inputenc}             % allow input of utf-8 encoded characters
\usepackage{crimson}                    % document's serif font, in the style of minion pro
\usepackage{helvet}                     % document's sans serif font, helvetica

\usepackage[strict,autostyle]{csquotes} % smart and nestable quote marks
\usepackage[USenglish]{babel}           % automatically regionalize hyphens, quote marks, etc
\usepackage{microtype}                  % improves text appearance with kerning, etc

\usepackage{abstract}                   % allow full-page title/abstract in twocolumn mode
\usepackage{authblk}                    % footnote-style author/affiliation info
\usepackage{booktabs}                   % better looking tables
\usepackage{caption}                    % custom figure/table caption styles
\usepackage[final]{draftwatermark}      % watermark paper as a draft
\usepackage{geometry}                   % configure page dimensions and margins
\usepackage{graphicx}                   % better inclusion of graphics
\usepackage{hyperref}                   % hypertext in document
\usepackage{rotating}                   % rotate wide tables or figures on a page to make them landscape
\usepackage{setspace}                   % configure spacing between lines
\usepackage{titlesec}                   % custom section and subsection heading
\usepackage{url}                        % make nice line-breakble urls
\usepackage[numbers]{natbib}    % author-year citations w/ bibtex, including textual and parenthetical
\usepackage{adjustbox}
\usepackage{multirow}
\usepackage{subfigure}

\usepackage{amsmath}
\usepackage{algorithm}
\usepackage[noend]{algpseudocode}

\usepackage{color}

\makeatletter
\def\BState{\State\hskip-\ALG@thistlm}
\makeatother


\titleformat{\section}{\normalfont\sffamily\large\bfseries\color{black}}{\thesection.}{0.3em}{}
\titleformat{\subsection}{\normalfont\sffamily\small\bfseries\color{black}}{\thesubsection.}{0.3em}{}

\hypersetup{
	pdfauthor={\myname},
	pdftitle={\papertitle},
	pdfsubject={\papertitle},
	pdfkeywords={\paperkeywords},
	pdffitwindow=true,         % window fit to page when opened
	breaklinks=true,           % break links that overflow horizontally
	colorlinks=false,          % remove link color
	pdfborder={0 0 0}          % remove link border
}

\SetWatermarkText{DRAFT}
\SetWatermarkScale{1.3}
\SetWatermarkLightness{0.9}

\begin{document}

\title{\papertitle}
\date{\paperdate}
\author{\myname$^1$, Federico Battiston$^{1}$, Gerardo Iñiguez$^{1,2,3}$, Michael Szell$^{4,5}$\footnote{\myemail}}

\affil[]{\small
					$1$) \myaffiliation\\
					$2$) Department of Computer Science, Aalto University School of Science, FI-00076 Aalto, Finland\\
					$3$) IIMAS, Universidad Nacional Auton\'{o}ma de M\'{e}xico, 01000 Ciudad de M\'{e}xico, Mexico\\
					$4$) %NEtwoRks, Data, and Society (NERDS), 
					NEtwoRks, Data, and Society (NERDS), IT University of Copenhagen, DK-2300 Copenhagen, Denmark \\
					$5$) Complexity Science Hub Vienna, 1080 Vienna, Austria\\
				}

% make title and abstract full-page single-column
%\twocolumn[
\maketitle
\begin{onecolabstract}
Urban transportation networks, from sidewalks and bicycle paths to streets and rail lines, provide the backbone for movement and socioeconomic life in cities. These networks can be understood as layers of a larger multiplex transport network. Because most cities are car-centric, the most developed layer is typically the street layer, while other layers can be highly disconnected. To make urban transport sustainable, cities are increasingly investing to develop their bicycle networks. However, given the usually patchy nature of the bicycle network layer, it is yet unclear how to extend it comprehensively and effectively given a limited budget. Here we develop data-driven, algorithmic network growth strategies and apply them to cities around the world, showing that small but focused investments allow to significantly increase the connectedness and directness of urban bicycle networks. We motivate the development of our algorithms with a network component analysis and with multimodal urban fingerprints that reveal different classes of cities depending on the connectedness between different network layers. We introduce two greedy algorithms to add the most critical missing links in the bicycle layer: The first algorithm connects the two largest connected components, the second algorithm connects the largest with the closest component. We show that these algorithms outmatch both a random approach and a baseline minimum investment strategy that connects the closest components ignoring size. Our computational approach outlines novel pathways from car-centric towards sustainable cities by taking advantage of urban data available on a city-wide scale. It is a first step towards a quantitative consolidation of bicycle infrastructure development that can become valuable for urban planners and stakeholders.


\vspace{0.5cm}
\end{onecolabstract}%]
%\saythanks % typeset any footnotes arising in the above block

% Blocked comments
\iffalse
STORY LOGIC (targeted journal: R Soc Interface)

1. INTRO

2. DATA AND NETWORK CONSTRUCTION

3. CLUSTERING CITIES THROUGH THE OVERLAP CENSUS
- want to understand multimodal potential with overlap census
- this allows us to classify cities

4. GROWING THE BICYCLE NETWORK
- we notice: the bike layer is sometimes really patchy
- but good bike infra needs connectedness and directness
- therefore, lets connect it. also measure directness

5. DISCUSSION
- Just first step. Need more on directness or to see how to bring together locally applied strategies ( fromlocal district budgets/planning) etc

\fi

\pagebreak

\section{Introduction}
%Start with the main problem: Street vs other (bike) networks
Most modern cities have followed a car-centric development in the 20th century \cite{Jacobs1961} and are today allocating a privileged amount of urban space to automobile traffic \cite{Gossling2016,Szell2018}. From a network perspective, this space can be described as the street layer of a larger mathematical object, the multiplex transport network \cite{Morris2012,Strano2015,Aleta2017}. A city's multiplex transport network contains other network layers that have co-evolved with the street layer, such as the bicycle layer or the rail network layer, which together constitute the multimodal transportation backbone of a city. Due to the car-centric development of most cities, street layers are the most developed layers and define or strongly limit other layers: For example, sidewalks are by definition footpaths along the side of a street and make up a substantial part of a city's pedestrian space \cite{Gossling2016}; similarly most bicycle paths are part of a street or are built along the side. From an urban sustainability perspective, this situation is suboptimal because the unsustainable mode of automobile transportation dominates sustainable modes like cycling. Consequently, urban planning movements in a number of pioneering cities are increasingly experimenting with drastic policies, such as applying congestion charges (London) \cite{Eliasson2008} and repurposing or removing car parking (Amsterdam, Oslo) \cite{Littke2016,bliss2019hcp,Nieuwenhuijsen2016}. These scattered efforts have shown preliminary success, however, a quantitative framework for  developing and assessing systematic strategies is missing.

% What we do, and why specific bike network
Here we consider the transport networks of 15 world cities and develop an urban fingerprinting technique based on multiplex network theory to characterize the various ways their transport layers are interconnected, outlining the potential for multimodal transport. Using clustering algorithms on the resulting urban fingerprints, we find clear classes of cities reflecting their transport priorities. We uncover network fragmentation within different layers and find that the bicycle layer is the most fragmented mobility infrastructure. To improve a city's vital dedicated bicycle infrastructure \cite{ Dill2013,Schoner2014,Hull2014,Buehler2016}, we develop algorithms for connecting disconnected graphs based on concrete quality metrics from bicycle network planning \cite{HannahTwaddellICF2018} and apply them to the empirical bicycle networks via network growth simulations. We find that localized investment into targeted missing links can rapidly consolidate fragmented bicycle networks, allowing to significantly increase their connectedness and directness, with potentially crucial implications for sustainable transport policy planning.

\section{Data acquisition and network construction}
We acquired urban transportation networks from multiple cities around the world using OSMnx  \cite{Boeing2017a}, a Python library to download and construct networks from OpenStreetMap (OSM). These data sets are of high quality \cite{Haklay2010b,Girres2010} in terms of correspondence with municipal open data \cite{Ferster2019} and completeness: More than $80\%$ of the world is covered by OSM \cite{Barbosa-Filho2017}. In particular, OSM's bicycle layer has better coverage than proprietary alternatives like Google Maps \cite{Hochmair2012}. We collect data from a diverse set of cities to capture different development states of bicycle infrastructure networks; from consolidated networks like Amsterdam and Copenhagen, less developed ones like Manhattan and Mexico City, to rapidly developing cities like Jakarta and Singapore. The various analyzed urban areas and their properties are reported in Table~\ref{tab:Table1} in the Supplementary Information (SI). Figure \ref{fig:fig01} shows the different network layers for two of our analyzed cities. The pedestrian layer contains all sidewalks and pedestrian paths. The bicycle layer consists of all infrastructure exclusive to bicycles, like designated cycleways and bicycle trails. The rail network captures all public transportation that uses rails, including subways and tramways. Finally the street layer is composed of all streets designated for motor vehicles.

The data to replicate the results can be downloaded from Harvard Dataverse (https://doi.org/10.7910/DVN/GSOPCK), the code is available as Jupyter Notebooks (https://github.com/nateraluis/bicycle-network-growth).

We characterize each city as a multiplex network \cite{Boccaletti2014,Kivela2014,Battiston2017a} with $M$ layers and $N$ nodes that can be active in one or more layers in the system. Layers are represented by a primal approach \cite{Porta2006} in which nodes are intersections (that may be present in one or more layers), while links represent streets (s), bicycle paths and designated bicycle infrastructure (b), subways, trams and rail infrastructure (r), or pedestrian infrastructure (p). This recent approach has been useful to demonstrate how cities grow \cite{Strano2012,Barthelemy2013}, how efficient \cite{Gallotti2014} and dense they are, and to capture the tendency of travel routes to gravitate towards city centers \cite{Lee2017}. Each layer $\alpha=1, \dots, M$ is described by an adjacency matrix $A^{[\alpha]}=\{a_{ij}^{[\alpha]}\}$ where $a_{ij}^{[\alpha]}=1$ if there is a link between nodes $i$ and $j$ in layer $\alpha$ and 0 otherwise. The multiplex urban system is then specified as a vector of adjacency matrices $\textbf{A} = (A^{[1]}, \ldots, A^{[M]})$.

\begin{figure*}[t!]
	\centering
	\includegraphics[width=\textwidth]{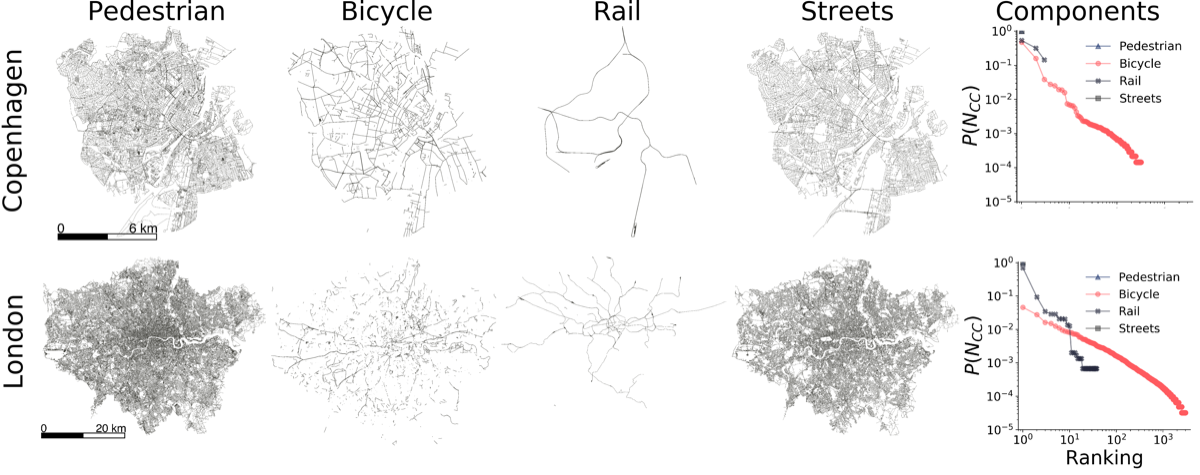}
	\caption{{\bf (Map plots, left)} Networks representing various layers of transport infrastructure (pedestrian paths, bicycle paths, rail lines, and streets) for  Copenhagen and London, with data from OpenStreetMap. {\bf (Right)} Connected component size distribution $P(N_{cc})$ as a function of the ranking of the component for all considered network layers and cities. All layers are well connected except the bicycle layer: Copenhagen has 321 bicycle network components despite being known as a bicycle-friendly city, while London's bicycle layer is much more fragmented, featuring over 3000 disconnected components. Copenhagen's largest connected bicycle component (leftmost data point) spans 50\% of the network, but London's only less than 5\%.}
	\label{fig:fig01}
\end{figure*}

\section{Quantifying multimodality with the overlap census}

\begin{figure*}[t!]
	\centering
	\includegraphics[width=0.8\textwidth]{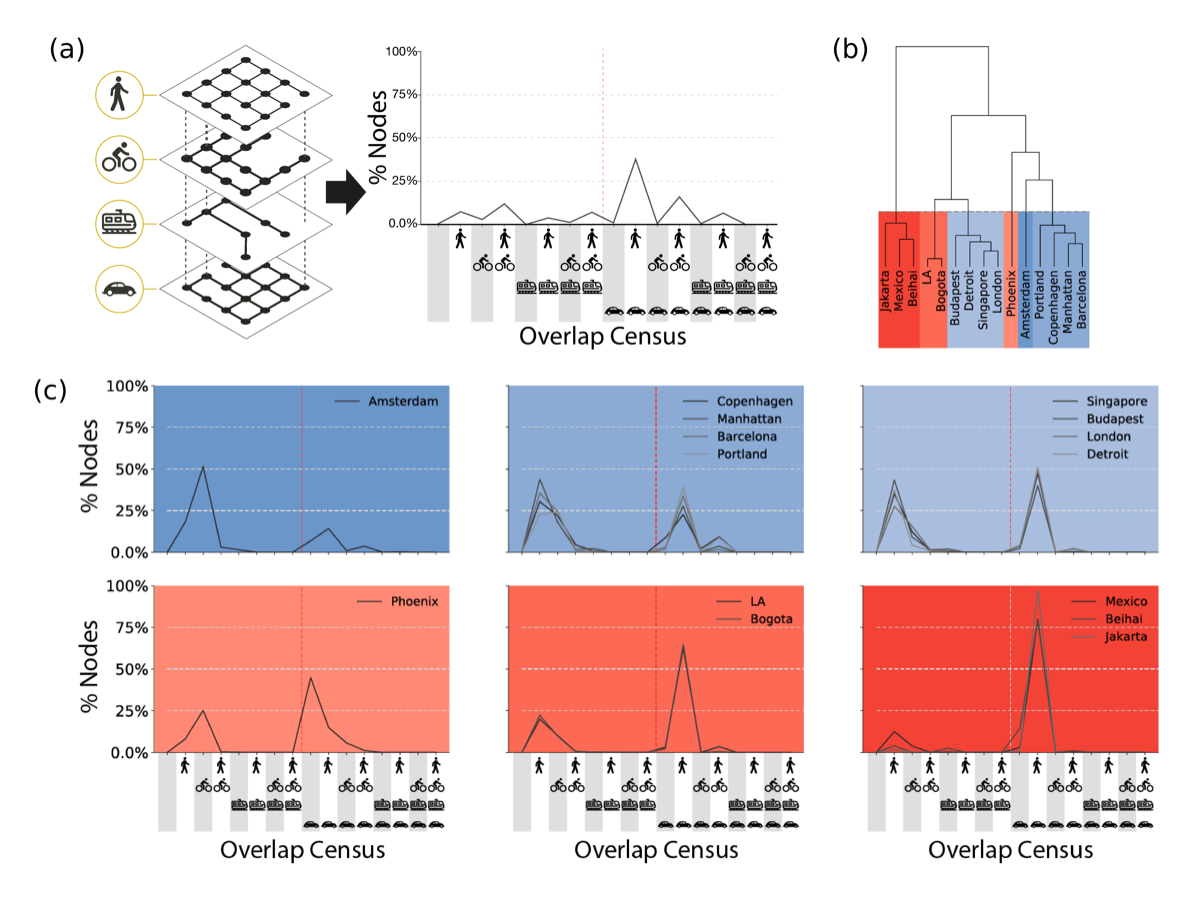}
	\caption{
		\textbf{(a)} Schematic of multiplex layers in a city (left) and its transformation to the overlap census (right). In the overlap census, the vertical red line gives a visual separation of the left from the right half where nodes become active in the street layer. High spikes in the right half indicate car-centricity.
		 \textbf{(b)} Clusters of cities based on similarity of their overlap census. We find six different clusters using a k-means algorithm (coloured areas), which explain more than $90\%$ of the variance.
		 \textbf{(c)} Overlap census for cities in each cluster. The first one corresponds to Amsterdam (the city with most active nodes in bicycle-only configurations). The Copenhagen-Manhattan-Barcelona-Portland city cluster has many active nodes in pedestrian-only and bicycle-only configurations, representing an active mobility city. The clusters of Los Angeles-Bogota and Mexico-Beihai-Jakarta are car-centric.}
	\label{fig:fig02}
\end{figure*}

% Why the overlap census is needed
Whereas one of the simplest features of single layer networks is the degree distribution, in multiplex networks a node can have different degrees in each layer, which inform us about the multimodal potential of a city through the different roles that its intersections play. If a city has nodes that are mainly active in one layer but not in others, there is no potential for multimodality. On the contrary, in a multimodal city we expect to find many transport hubs that connect different layers, such as train stations with bicycle and street access, i.e. nodes that are active in different multiplex configurations. Note that even in a multimodally ''optimal'' city there will be a high heterogeneity of node activities due to the different speeds and nature of transport modes, implying, for example, a much lower density of nodes necessary for a train network than for a bicycle network. Still, if we had a way to see and compare all combinations of node activities in the system, we could learn how much focus a city puts on connecting different modes. We can define such a fingerprint using the multiplex network formalism. We call the plot that counts the combinations of node activities a city's \emph{overlap census} (Figure \ref{fig:fig02}). Similar to edge overlap \cite{Bianconi2013,Battiston2014a} and multiplex motifs \cite{Battiston2017} that provide a characterization of multiplexity at the local scale, the overlap census captures the percentage of nodes that are active in different multiplex configurations and provides an "urban fingerprint'' of multimodality \cite{Aleta2017}.

To define the overlap census formally, we calculate the degree of each node in layer $\alpha$ as $ k_i^{[\alpha]}=\sum_ja_{ij}^{[\alpha]} $. We store degrees in a vector for each layer, \textbf{\textit{k}}$^{[\alpha]}=(k_i^{[\alpha]}, \dots, k_n^{[\alpha]})$, indicating layers by `p' for pedestrian, `b' for bicycle, `r' for rail infrastructure, and `s' for streets, as before. The use of vectorial variables like \textbf{A} and \textbf{\textit{k}}$^{[\alpha]}$ instead of those of the aggregated network lets us capture the richness of the system and work in the various layers independently. In Fig.~\ref{fig:fig02}(a) we show a schematic of how the overlap census is built: taking the multiplex network, transforming it to its corresponding degree vectors for all layers in the system, and calculating the percentage of nodes that overlaps in different configurations. Note that the overlap census provides more information than a simple counting of nodes, edges, or other single-layer network measures. The multiplex approach addresses the multimodality of a city: it not only counts how many nodes or links there are in each layer, but it shows how they are combined, revealing the possible multimodal mobility combinations in the city. Understanding the possibilities for interchange between mobility layers provide us with a better understanding of urban systems, showing us the complexity and interplay between layers.

% What we can do with the overlap census: classification
Due to the expected heterogeneity of node activities in different layers, the overlap census of a specific city is also expected to be heterogeneous and hard to assess on its own. Therefore, a good way to assess a city's overlap census is by comparing it with the overlap census of other cities. Explicitly, we find similarities between cities via a k-means algorithm. The algorithm separates the 15 analyzed cities into six different clusters [Fig.~\ref{fig:fig02}(b)]. On the left half of the overlap census we show the configurations in which nodes are not active in the street layer, while the right half contains car-related configurations. These clusters of cities are useful to explain similarities in infrastructure planning in different urban transport development paths \cite{Rodrigue2013,Louf2014}, with clusters of car-centric urbanization (like Mexico, Beihai, and Jakarta) opposed to clusters that show a more multimodally focused evolution of their urban mobility infrastructure (like Copenhagen, Manhattan, Barcelona, and Portland). In the extreme cluster that contains only Amsterdam, close to $50\%$ of nodes are active in the bicycle layer, while in the Mexico-Beihai-Jakarta cluster more than $50\%$ of nodes are active in the street-pedestrian configuration. This concentration of nodes in just one configuration informs us not only about the (sometimes already well-known) mobility character of the city, i.e. Amsterdam being a bicycle-friendly city, but unveils the importance of explicitly considering overlooked layers and their interconnections. For example, Singapore, Budapest, London, and Detroit have two main peaks indicating that most of their nodes are either active in the street-pedestrian or only in the pedestrian configuration, i.e. there are plenty of walkable areas exclusive to pedestrians. This is not the case in Los Angeles and Bogota, where the majority of nodes is active in the car-pedestrian combination, i.e. the pedestrians have to share most of the city with cars.
%While the cases of Amsterdam, and the car-centric cities are extreme, other cases as the one of Copenhagen, Manhattan, Barcelona and Portland, show overlap profiles with a more heterogeniously distributed. Or the case of Singapore, Budapest, London and Detroit with two main peaks, indicating that either the nodes are active in the street-pedestrian or in the only pedestrian configuration. 

\section{Defining bicycle network growth strategies and quality metrics}
% State the prpoblem: patchy infra, which is bad according to lit
Across all cities considered, the data reveal that the pedestrian and street layers are the most connected, while the bicycle and rail layers are the most fragmented, particularly the former (see Table \ref{tab:Table1} in SI). The overlap census reflects this fact, as configurations where bicycles are active are less frequent than configurations where they are not active. To quantify such an underdevelopment in the sustainable mobility infrastructure, we focus on the single layer of bicycle networks and on two well-established metrics in bicycle infrastructure quality assessment \cite{Krizek2005,movement2013,Dobrovolny2014,HannahTwaddellICF2018,Beck2019}: {\it connectedness} and {\it directness}. Connectedness indicates ``the ease with which people can travel across the transportation system'' \cite{HannahTwaddellICF2018}, and it is related to answering the question ``can I go where I want to, safely?''. Directness addresses the question ``how far out of their way do users have to travel to find a facility they can or want to use?'', and can be measured by how easy it is to go from one point to another in a city using bicycle infrastructure versus other mobility options, like car travel.

We choose to measure connectedness and directness over the designated bicycle infrastructure only, without considering travel on streets. Although it is possible to cycle on streets, growing evidence from bicycle infrastructure and safety research is unveiling serious safety issues for cycling when mixed with vehicular traffic \cite{Reynolds2009,Teschke2012,Pucher2016}. To quantify connectedness, we first measure the number of disconnected components of each city's bicycle network. It is no surprise that car-centric cities have a highly fragmented bicycle infrastructure: for example, London has more than 3,000 disconnected bicycle infrastructure segments. However, even bicycle-friendly cities like Copenhagen have over 300 disconnected bicycle path components -- see the connected component size distribution $P( N_{cc} )$ in Fig.~\ref{fig:fig01}. This infrastructure fragmentation in the bicycle layer poses a challenge for a city's multimodal mobility options and for the safety of its cycling citizens \cite{Dill2009,Chataway2014}.

% How people tried to solve it
There are various approaches in developing automated strategies for bicycle infrastructure planning. Hyodo et al.~\cite{Hyodo2000} have proposed a bicycle route choice model to plan bicycle lanes taking into account facility characteristics. Other studies have used input data from bicycle share systems \cite{Bao2017} or origin destination matrices \cite{Mauttone2017} to plan bicycle lanes. More recently, taxi trips have been used to identify susceptible clusters for bicycle infrastructure \cite{Akbarzadeh2018a}.  Here we attempt an alternative approach: Since hundreds of bicycle network components already exist in most cities, we aim at consolidating the existing infrastructure by making strategic connections between components rather than starting from scratch. %Obviously this approach is not applicable to cities that have no bicycle infrastructure.  %It could be much easier to improve by finding and connecting the ``missing links'' than by starting from zero.

%This result is coherent with the data reported in Table \ref{tab:Table1}, where we saw that the bicycle layer is the most fragmented one, with some cities like London that have more than 3,000 different bicycle infrastructure disconnected segments. Good bicycle infrastructure is all about 1) connectedness, and 2) directness , these two main characheristics are important to promote the use of bicycles as a transportation alternative. But with thousands of components, how to connect well?

%In the supplementary information Table \ref{tab:Table1} we report the number of connected components and nodes in each one of the layers, for most of the cities with most of the cities having the driving and pedestrian layers consolidated in one giant component respectively, while the rail and bicycle layer are fragmented in multiple components (e.g. Copenhagen bike layer has 321 components, while London has 3,023). The infrastructure fragmentation, especially in the bicycle layer, poses a challenge for the multimodality mobility options as well as for the users safety \cite{Dill2009,Chataway2014}. For most cities the number of components in the bicycle layer is around a few hundreds, the only outlier is London with more than one thousand components.

% How we do it better/differently
Our approach takes into account the currently available bicycle infrastructure and uses an algorithmic process to improve the network by finding the most important missing links step by step. This way we focus on optimizing the connectedness metric, growing the bicycle infrastructure by making it more connected, merging parts into fewer and fewer components. We develop two iterative greedy algorithms that we check against a random and a minimum investment approach. The first algorithm, \emph{Largest-to-Second} (L2S), identifies in each step the largest connected component and connects it to the second largest. The second algorithm, \emph{Largest-to-Closest} (L2C), also identifies the largest connected component, but connects it to the closest of the remaining components. See the SI for details. In both algorithms, components are connected through a direct link between their two closest nodes. We use this technique as an approximation to the underlying street-shortest path -- since the most relevant shortest 100 connections typically range from $14$ to $500$ meters, roughly the length of two blocks, this approximation is reasonable. The algorithms repeat this process until there are no more disconnected components in the network. 

\begin{figure*}[htbp]
	\centering
	\includegraphics[width=0.8\textwidth]{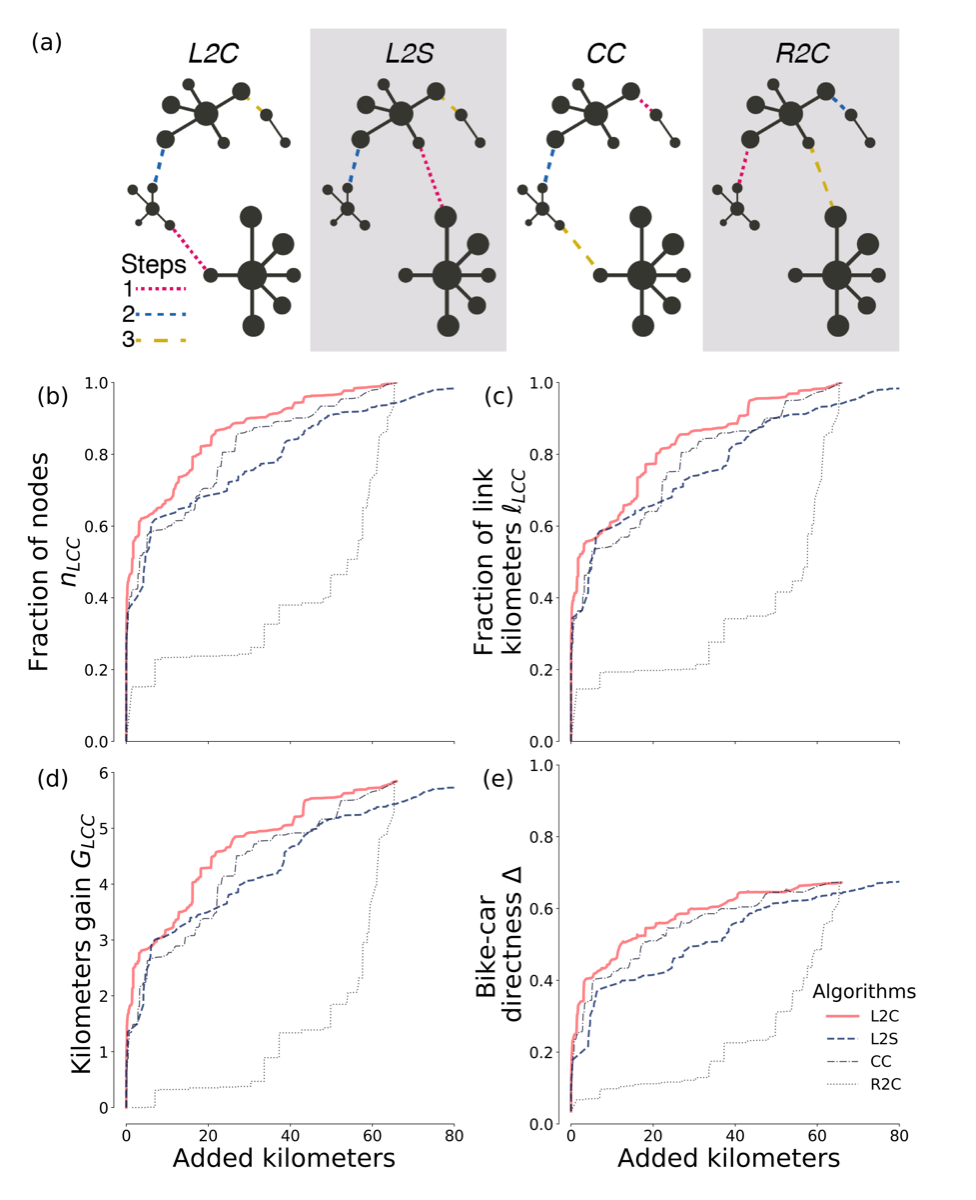}
	\caption{\textbf{(a)} Schematic representation of algorithms to improve bicycle network infrastructure: Largest-to-Closest (L2C) finds the largest component and connects it with the closest one; Largest-to-Second (L2S) connects the largest component with the second largest; Closest-Connected (CC) connects the two closest components; and Random-to-Closest (R2C) picks a random component and connects it to the closest. \textbf{(b)} Normalized increase in nodes inside the largest connected component ($n_{LCC}$). \textbf{(c)} Normalized increase in kilometers inside the largest connected component ($\ell_{LCC}$). \textbf{(d)} Kilometers gain ($G_{LCC}$). \textbf{(e)} Bicycle-car directness ($\Delta$). Measures in (b-e) are plotted as a function of the sum of added links in kilometers, for the case of Budapest.}
	\label{fig:fig03}
\end{figure*}

To have a random baseline, we compare our algorithms with a \emph{Random-to-Closest} (R2C) component approach. In each step of this baseline approach, one component is picked at random and connected with the closest remaining one. This baseline allows us to model a scenario where infrastructure is developed following a systematic but random linking approach -- in urban development this corresponds to uncoordinated local planning that randomly connects close pieces of bicycle infrastructure. We also implement a second baseline, the extreme case of \emph{Closest-Components} (CC), which prioritizes connecting the closest two components disregarding their size. This CC approach is equivalent to an ``invest as little as possible'' development strategy -- it builds up a minimum-spanning-tree-like structure. All four algorithms connect components optimizing a well-defined criterion, finding the critical missing links in the network, and adding one new link per iteration. See Fig.~\ref{fig:fig03}(a) for a schematic of the four algorithms.

To test how much cities improve their bicycle layers using these four algorithms, we define two metrics on the bicycle layer that operationalize the notion of connectedness: i) $n_{LCC} = \frac{N_{LCC}}{N} $, the fraction of nodes inside the largest connected component ($N_{LCC}$) compared to the total number of nodes ($N$), and ii) $\ell_{LCC} =  \frac{L_{LCC}}{L} $, the fraction of link kilometers inside the largest connected component ($L_{LCC}$) compared to the total number of link kilometers in the network ($L$). Both metrics take values between 0 and 1, where 1 means that there is only one connected component. An intermediate value, for example 0.2, means that the largest connected component contains 20\% of all bicycle intersections or path kilometers. Executing our algorithms step by step these metrics can only grow, approaching 1 when the process is complete and they terminate. What distinguishes the algorithms is \emph{how fast} these values grow.

We quantify directness through the metric of: iii) bicycle-car directness $\Delta$, which answers the question ``how direct are the average routes of bicycles compared to cars?'' For example, if the shortest car-route from west to east Manhattan is 4\,km instead of the straight-line distance of 3\,km, then the car's route has a (car) directness of $0.75$. If the shortest route on the bicycle network between these two points is 5\,km, the bicycle's route has a (bicycle) directness of $0.6$. Comparing these two values yields a bicycle-car directness of $0.6/0.75 = 0.8$ for this route. The bicycle-car directness $\Delta$ averages this value over all possible routes. Note that if the bicycle network is a subset of the street network, then $\Delta$ cannot be larger than $1$. Formally we write $\Delta=\frac{\langle\delta_{ij}^b\rangle_{ij}}{\langle\delta_{ij}^s\rangle_{ij}}$, where $\langle\delta_{ij}^{\alpha}\rangle_{ij}=\frac{d_{ij}}{d_{ij}^{\alpha}}$ is the average directness in layer $\alpha$ over all nodes $i$ and $j$ existing in that layer, with $d_{ij}$ the euclidean distance between nodes $i$ and $j$, and $d_{ij}^{\alpha}$ the length of the shortest path between them in layer $\alpha$. In each iteration of any of our algorithms, we implement this measure by randomly selecting one thousand pairs of origin-destinations nodes and then averaging the corresponding street/bicycle directness (for the bicycle layer, a route between disconnected components has directness 0).

Finally, in order to measure the cumulative efficiency of our algorithms, we define the metric: iv) $G_{LCC}$ as the relative gain of bicycle path kilometers in the largest connected component. For example, $G_{LCC}=1.5$ means that the algorithm has increased the largest connected component's original size by 150\%. Formally, $G_{LCC}=\frac{L_{LCC}-L_{LCC_0}}{L_{LCC_0}}$, where $L_{LCC_0}$ is the sum of kilometers in the largest connected component before the algorithm runs. As with all other metrics, $G_{LCC}$ is monotonically increasing with the growth algorithm, and reaches $\frac{1-\ell_{LCC_0}}{\ell_{LCC_0}}$ at the end of the dynamics.

\section{Growing bicycle networks}
We demonstrate in Fig.~\ref{fig:fig03} the power of the various growth strategies by showing the initial state of the bicycle layer for the case of Budapest and its state after 85 iterations of the \emph{Largest-to-Closest} algorithm: At this point the network has almost quadrupled the size of its largest connected component (from 82\,km to 313\,km), with a negligible investment of just less than 5\,km in new connecting bicycle paths. In terms of connectedness, it goes from $15\%$ to $56\%$ connected. This rapid increase shows that the city can easily improve its bicycle infrastructure with small investments. For some extreme cases, like Bogota, with the same 5\,km investment, the bicycle-car directness increases from $6\%$ to almost $48\%$ and connectedness from $34\%$ to $89\%$. Similar encouraging results hold for other cities (see SI). 

The fraction of nodes inside the largest connected component increases rapidly with newly added links for all considered algorithms except \emph{Random-to-Closest}, Fig.~\ref{fig:fig03}(b). The \emph{Largest-to-Closest} algorithm performs better than the others, even more than \emph{Closest-Components} which prioritizes minimum investments in the network. Since we are considering bicycle infrastructure, a better practical measure than the number of intersections is the number of kilometers that can be cycled using only designated paths. Figure~\ref{fig:fig03}(c) shows how this measure improves in a similarly explosive way: with an investment of only 20\,km, the largest connected component will contain $~80\%$ of the original bicycle infrastructure. Results for the kilometer gain $G_{LCC}$ are shown in Fig.~\ref{fig:fig03}(d). Three of the four algorithms rapidly gain new kilometers, but as the invested new kilometers grow, each algorithm follows a different gain rate. Also for this metric, \textit{Largest-to-Closest} is the algorithm with the best performance.

We also measure the bicycle-car directness ratio, Fig.~\ref{fig:fig03}(e). The bicycle-car directness $\Delta$ improves as the algorithms consolidate the network. These improvements are, however, indirectly driven by the improvement of connectedness, which boosts the accessibility of bicycles to different areas of the city. The flattening of the curves at a value considerably smaller than 1 (around 0.65) shows that cars will always outperform bicycles in terms of directness, having on average at least 33\% shorter paths in the city. This suboptimal flattening is a natural consequence of the algorithms optimizing for connectedness only, not adding ``redundant'' connections. Nevertheless, the measure shows that, similar to connectedness, with a relatively negligible investment of bicycle path kilometers into the system, the bicycle network's directness improves drastically, even in the greediest case where the shortest possible missing link is added in every iteration. This result holds for all analyzed cities (see SI). The large differences between the baseline \emph{Random-to-Closest} and our two algorithms (\emph{Largest-to-Second} and \emph{Largest-to-Closest}) show the importance of following an approach that consolidates and grows the largest connected component.

\begin{figure*}[t!]
	\centering
	\includegraphics[width=0.67\textwidth]{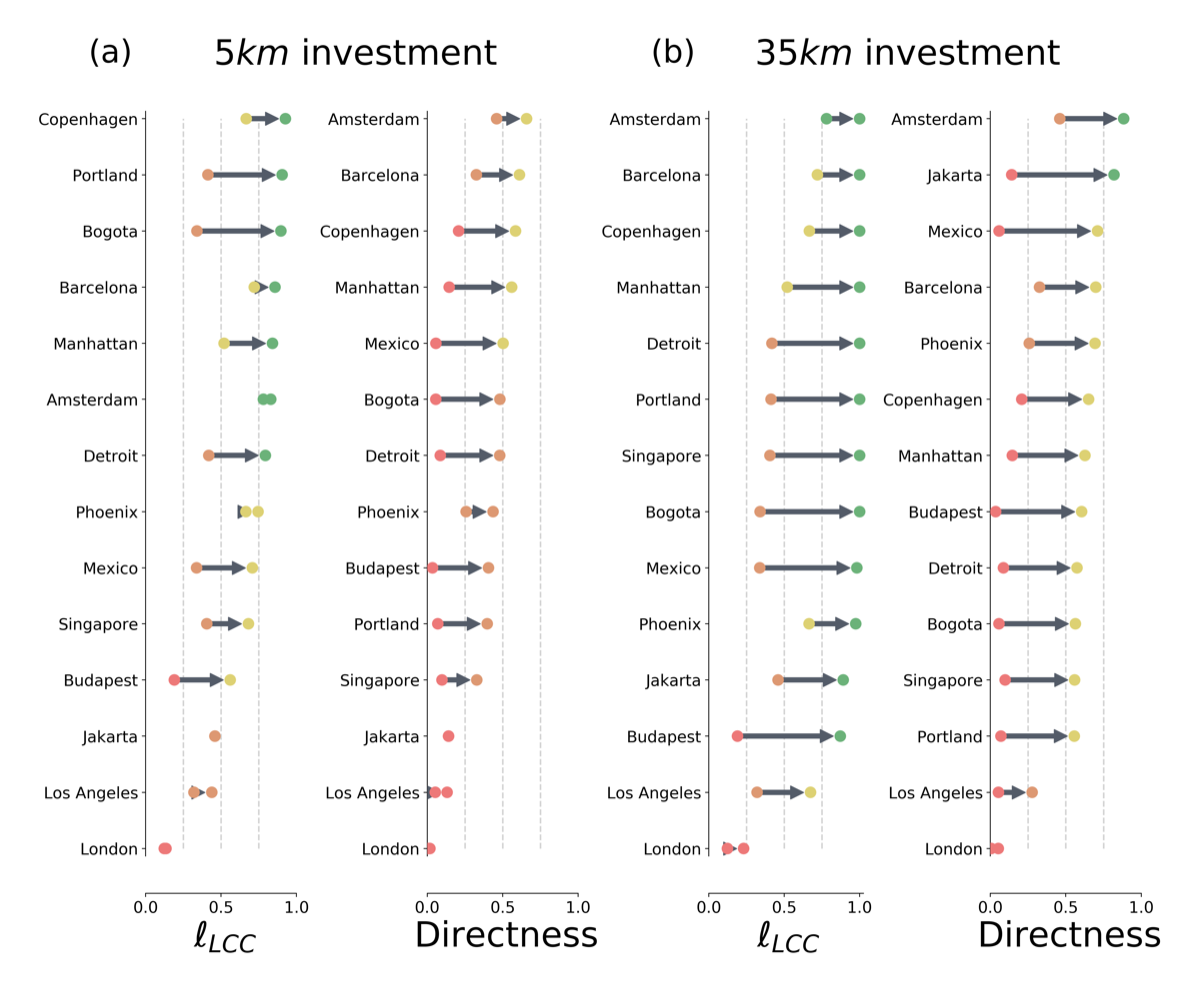}
	\caption{Cities improvement and ranking using the \emph{Largest-to-Closest} algorithm. We report the improvement and ranking on the fraction of total kilometers of bicycle infrastructure in the largest connected component ($\ell_{LCC}$) and in the bicycle-car directness ($\Delta$). Dotted lines show thresholds of $25\%$, $50\%$, and $75\%$. Plots (a-b) show investment strategies of 5\,km and 35\,km, respectively.}
	\label{fig:fig04}
\end{figure*}

Since every city has a unique overlap census, Fig.~\ref{fig:fig02}(c), differences also arise in the state of the bicycle layer and its improvement after applying a growth algorithm. To see this effect, we rank how cities improve using the \textit{Largest-to-Closest} algorithm in two different investment scenarios: investing either i) 5\,km, or ii) 30\,km. Figure~\ref{fig:fig04}(a) shows how cities improve when investing 5\,km of bicycle infrastructure. We see that some cities get above $75\%$ of their existing infrastructure connected, meaning that their bicycle layer only needs a small extension. On the other hand, cities like London, Los Angeles, and Jakarta need a larger investment to improve. Concerning bicycle-car directness, cities reach lower values due to the focus of the algorithms on completeness. In the worst performing cities like Los Angeles, a covered length close to $50\%$ can be reached easily, while the bicycle-car directness ratio stays below $20\%$, showing that it is much harder to gain an acceptable bicycle infrastructure in cities where cars are overprioritized. The 35\,km investment strategy shows that most cities can get at least $75\%$ of their bicycle infrastructure connected, Fig.~\ref{fig:fig04}(b). The worst performing outlier is London, due to its bicycle layer containing more than 3000 connected components scattered around $1600$ km$^2$ (see Table \ref{tab:Table1}). In terms of bicycle-car directness London also performs badly, while Amsterdam is the best performing one.

\section{Discussion}
We have developed a transport multiplexity fingerprint for cities, and have shown that every city has a unique multiplex profile, with some cities displaying a car-centric profile and others more multimodal profiles. Independently of these profiles, one common characteristic of cities is the fragmentation of their bicycle layers. We have proposed the use of data-driven algorithms to consolidate bicycle network components into connected networks to improve efficiently sustainable transport. We have shown that connecting the bicycle infrastructure in an algorithmic way rapidly improves the connectedness and directness of the bicycle layer. These algorithms, when compared with two baselines, highlight the usefulness of growing the bicycle network on a city-wide scale (considering all areas of the city) rather than randomly adding local bicycle infrastructure. Improving the connectivity of bicycle lanes and paths improves not only the network itself, but also promotes the use of bicycles as means of transportation in a city, improving the health of its inhabitants \cite{Mueller2018}. 

%Limitations / future work
Improving bicycle infrastructure one link at the time (by identifying suitable components to connect) is only the first step towards a systematic framework for realistic bicycle network growth strategies. Our current approach is not the last word in this development, since it does not yet explicitly optimize for directness and does not account for transport flow. In these algorithms, each new link works as a bridge between components, potentially having large betweenness centrality. Such high-betweenness segments could become overused and create bottlenecks in practice. To improve this situation, it would be necessary to create links in the network that act as redundant paths. In doing so, directness would also be improved, along with the network's robustness to interruptions. This is an interesting and possibly demanding task that we leave for future research, as the new links would have to be created in a coherent manner balancing trade-offs between network structure and mobility dynamics. Despite these various possibilities for qualitative updates to the studied growth strategies, our first models have demonstrated the capability to generate substantial improvements with minimal effort.

% Last conclusion
The use of data-driven algorithms to identify crucially missing links in bicycle infrastructure has the potential to improve the mobility infrastructure of cities efficiently and economically. This approach is not only useful for planning city structure, but could also be used together with simulating mobility flows and to provide insights on how the system will behave after new measures are implemented. We anticipate that a future stream of work should include longitudinal studies in multiple cities, along with algorithmic simulations to first model and simulate possible changes to the transport network, and then to test those models with ground truth data, to compare the evolution of infrastructure and mobility dynamics between cities with different transport priorities.

\section*{Acknowledgments}
Icons in Figs.~\ref{fig:fig02}(a-b) and \ref{fig:fig03}(a) designed by Freepik for FlatIcon.com. The authors wish to thank Ana Paula Velasco and Roberta Sinatra for their comments and support.

% print the bibliography
\begin{small}
\setlength{\bibsep}{0.00cm plus 0.05cm} % no space between items
\bibliographystyle{unsrt}
\bibliography{library.bib}
\end{small}

\newpage
\section*{Supplementary information}
\renewcommand{\thesection}{S.I. \arabic{section}}
\renewcommand{\thetable}{S.I. \arabic{table}}
\renewcommand{\thefigure}{S.I.\arabic{figure}}
\setcounter{figure}{0}    
\setcounter{table}{0}

\subsection*{Data}
Table~\ref{tab:Table1} shows measures for the fifteen analyzed cities. For each layer in a city we report the number of nodes $N$ and the number of connected components $CC$.
\begin{table*}[h!]
	\centering
	\begin{adjustbox}{width=0.6\textwidth,keepaspectratio}
		\begin{tabular}{rr|rr|rr|rr|rr}
	\toprule
	 {} &{} & \multicolumn{2}{c}{Walk} & \multicolumn{2}{c}{Bike} & \multicolumn{2}{c}{Rail} & \multicolumn{2}{c}{Drive} \\
	{} & Area $km^2$ & $CC$ &       $N$ &    $CC$ &      $N$ &  $CC$ &     $N$ &  $CC$ &       $N$ \\
	\midrule
	Amsterdam  &221.77&  1 &   23,321 &   355 &  34,529 &   8 &  1,096 &   1 &   15,125 \\
	Barcelona  &104.80&  1 &   20,203 &   122 &   75,53 &  15 &   249 &   1 &   10,393 \\
	Beihai     &2,380&  1 &    2,026 &       0 &      0 &   3 &    59 &   1 &    2,192 \\
	Bogota     &614.45&  1 &   81,814 &   171 &   97,60 &  12 &   166 &   1 &   62,017 \\
	Budapest   &529.96&  1 &   73,172 &   257 &  10,494 &  20 &  1,588 &   1 &   37,012 \\
	Copenhagen &100.04&  1 &   30,746 &   321 &  13,980 &   3 &   276 &   1 &   15,822 \\
	Detroit    &431.43&  1 &   47,828 &    53 &   3,663 &   3 &    20 &   1 &   28,462 \\
	Jakarta    &5,679.87&  1 &  140,042 &    19 &    248 &   6 &    58 &   1 &  138,388 \\
	Los Angeles        &1,315.26&  1 &   89,543 &   230 &  14,577 &   9 &   173 &   1 &   71,091 \\
	London     &1,637.13&  1 &  270,659 &  3023 &  62,398 &  38 &  2988 &   1 &  179,782 \\
	Manhattan  &73.43&  1 &   13,326 &   105 &   3,871 &   5 &   349 &   1 &    5,671 \\
	Mexico     &1,489.42&  1 &  108,033 &    52 &   5,218 &  17 &   370 &   1 &   95,375 \\
	Phoenix    &1,348.43&  1 &  111,363 &   141 &  35,631 &   4 &   105 &   1 &   73,688 \\
	Portland   &382.22&  1 &   50,878 &   198 &  24,252 &   2 &   230 &   1 &   35,025 \\
	Singapore  &640.10&  1 &   82,808 &   104 &  12,981 &  14 &   683 &   1 &   50,403 \\
	\bottomrule
\end{tabular}
	\end{adjustbox}
	\caption{Measures for the administrative area of analyzed cities. The number of connected components ($CC$) and nodes ($N$) for each layer in all cities of our dataset are highly diverse due to the varying developmental levels and focus of transport. 
		\label{tab:Table1}}
\end{table*}

\begin{figure*}[h!]
	\centering
	\includegraphics[width=0.68\textwidth]{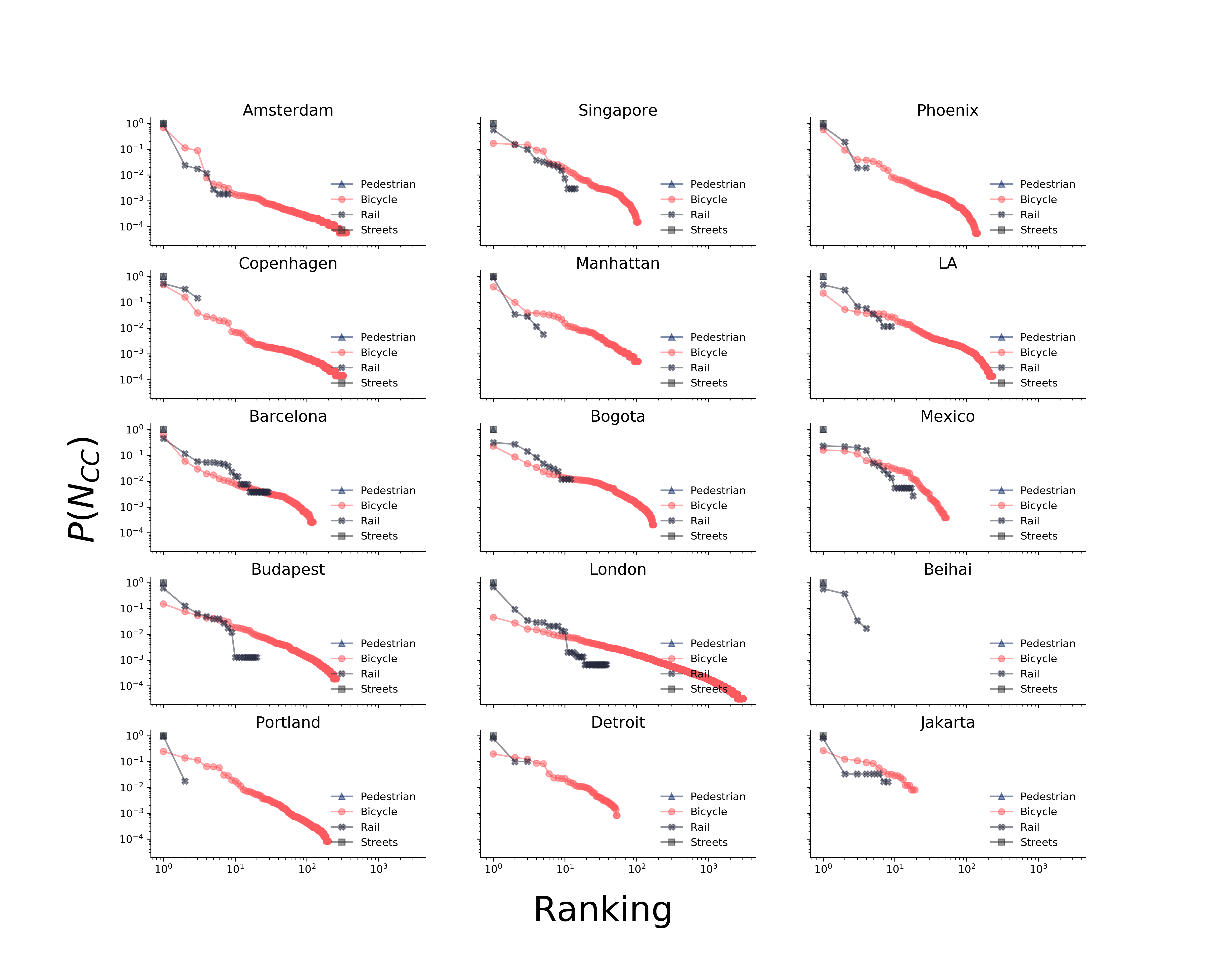}
	\caption{The connected component size distribution [$P(N_{cc})$] for all cities and layers is well connected except in the bicycle layer. London has the most fragmented bicycle infrastructure layer, with more than 3000 components}
	\label{fig:Nodes}
\end{figure*}

Figure \ref{fig:Nodes} shows the connected component size distribution $P(N_{cc})$ for all considered layers and cities.

\subsection*{Algorithms}
We develop two main algorithms to improve the bicycle layer. The first algorithm, \textit{Largest-to-Second}, identifies in each step the largest connected component and connects it to the second largest. The second algorithm, \textit{Largest-to-Closest}, also identifies the largest connected component, but in each step connects it to the closest of the remaining components.

To evaluate our algorithms we test them against a random baseline, \textit{Random-to-Closest}. In each step, \textit{Random-to-Closest} picks a component at random and connects it to the closest of the remaining components. We implement another baseline, the extreme case of \textit{Closest-Components}, which prioritizes connecting the closest two components.

\begin{algorithm}[h!]
	\caption{Largest-to-Second}\label{greedy_lcc}
	\begin{algorithmic}[1]
		\Procedure{\textit{L2S}}{}
		\State $\textit{G} \gets \text{ bicycle network graph}$
		\State $\textit{wcc} \gets \text{ components of network G}$
		\BState \emph{loop for n-1 components in wcc}:
		\State  \text{sort \textit{wcc} by components size}
		\State $\textit{cc} \gets \text{ two biggest components from \textit{wcc}}$
		\State $\textit{i\_j} \gets \text{ closest nodes between } cc_0 \text{ and } cc_1$
		\State $\text{connect } cc_0 \text{ and } cc_1$
		\State \textbf{goto} \emph{loop}.
		\BState \textbf{close};
		\EndProcedure
	\end{algorithmic}
\end{algorithm}

\begin{algorithm}[h!]
	\caption{Largest-to-Closest}\label{greedy_min}
	\begin{algorithmic}[1]
		\Procedure{\textit{L2C}}{}
		\State $\textit{G} \gets \text{ bicycle network graph}$
		\State $\textit{wcc} \gets \text{ components of network G}$
		\BState \emph{loop for n-1 components in wcc}:
		\State  \text{sort \textit{wcc} by components size}
		\State $cc_0 \gets \text{ biggest component from \textit{wcc}}$
		\State $cc_n \gets \text{ clossest component to } cc_0$
		\State $\textit{i\_j} \gets \text{ closest nodes between } cc_0 \text{ and } cc_n$
		\State $\text{connect } cc_0 \text{ and } cc_n$
		\State \textbf{goto} \emph{loop}.
		\BState \textbf{close};
		\EndProcedure
	\end{algorithmic}
\end{algorithm}

\begin{algorithm}[h!]
	\caption{Random-to-Closest}\label{random}
	\begin{algorithmic}[1]
		\Procedure{\textit{R2C}}{}
		\State $\textit{G} \gets \text{ bicycle network graph}$
		\State $\textit{wcc} \gets \text{ components of network G}$
		\BState \emph{loop for n-1 components in wcc}:
		\State $cc_{ran} \gets \text{ random component from \textit{wcc}}$
		\State $cc_n \gets \text{ clossest component to } cc_{ran}$
		\State $\textit{i\_j} \gets \text{ closest nodes between } cc_{ran} \text{ and } cc_n$
		\State $\text{connect } cc_{ran} \text{ and } cc_n$
		\State \textbf{goto} \emph{loop}.
		\BState \textbf{close};
		\EndProcedure
	\end{algorithmic}
\end{algorithm}

\begin{algorithm}[h!]
	\caption{Closest-Components}\label{min_delta}
	\begin{algorithmic}[1]
		\Procedure{\textit{CC}}{}
		\State $\textit{G} \gets \text{ bicycle network graph}$
		\State $\textit{wcc} \gets \text{ components of network G}$
		\BState \emph{loop for n-1 components in wcc}:
		\State $\Delta_{min} \gets \text{ clossest components in \textit{wcc}}$
		\State $cc_{0} \gets \text{ first component for }\Delta_{min}$
		\State $cc_{1} \gets \text{ second component for } \Delta_{min}$
		\State $\textit{i\_j} \gets \text{ closest nodes between } cc_{0} \text{ and } cc_{1}$
		\State $\text{connect } cc_{0} \text{ and } cc_{1}$
		\State \textbf{goto} \emph{loop}.
		\BState \textbf{close};
		\EndProcedure
	\end{algorithmic}
\end{algorithm}

\section*{Bicycle network improvement}

\begin{figure*}[h!]
	\centering
	\includegraphics[width=0.68\textwidth]{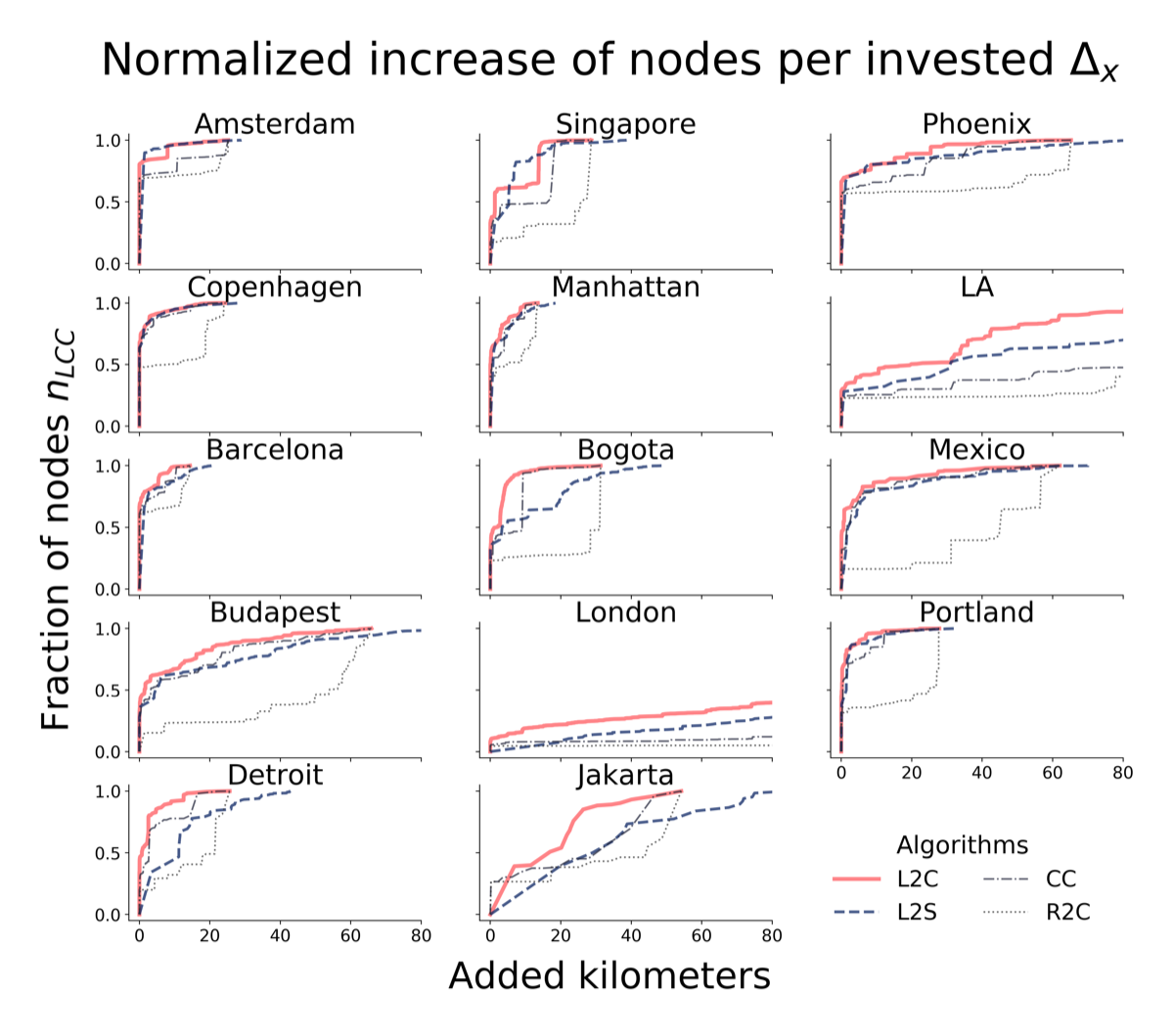}
	\caption{Normalized increase in nodes inside the largest connected component ($n_{LCC}$).}
	\label{fig:NodesIncrease}
\end{figure*}

\begin{figure*}[h!]
	\centering
	\includegraphics[width=0.68\textwidth]{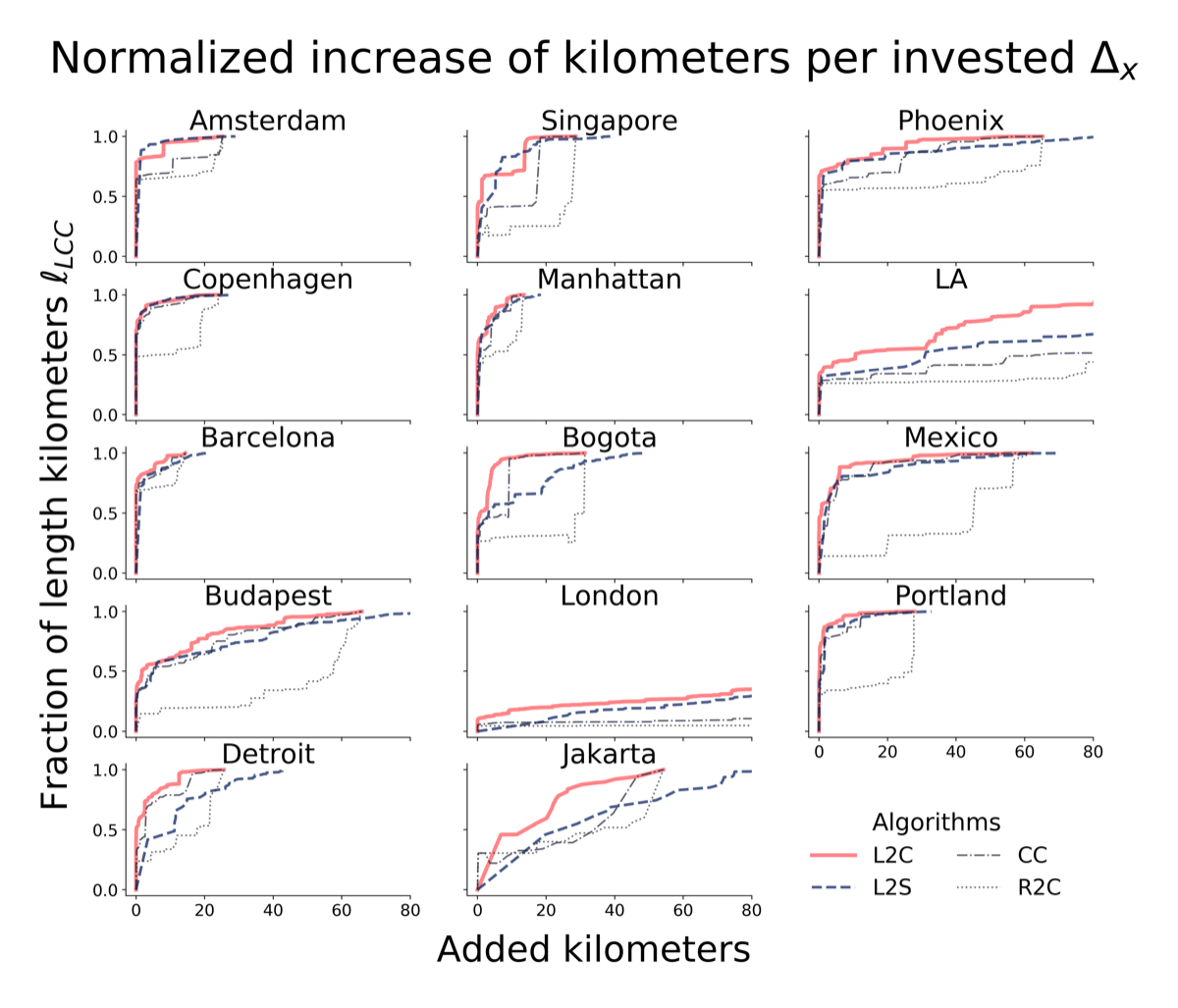}
	\caption{Normalized increase in kilometers inside the largest connected component ($\ell_{LCC}$).}
	\label{fig:Lengthsncrease}
\end{figure*}

\begin{figure*}[h!]
	\centering
	\includegraphics[width=0.68\textwidth]{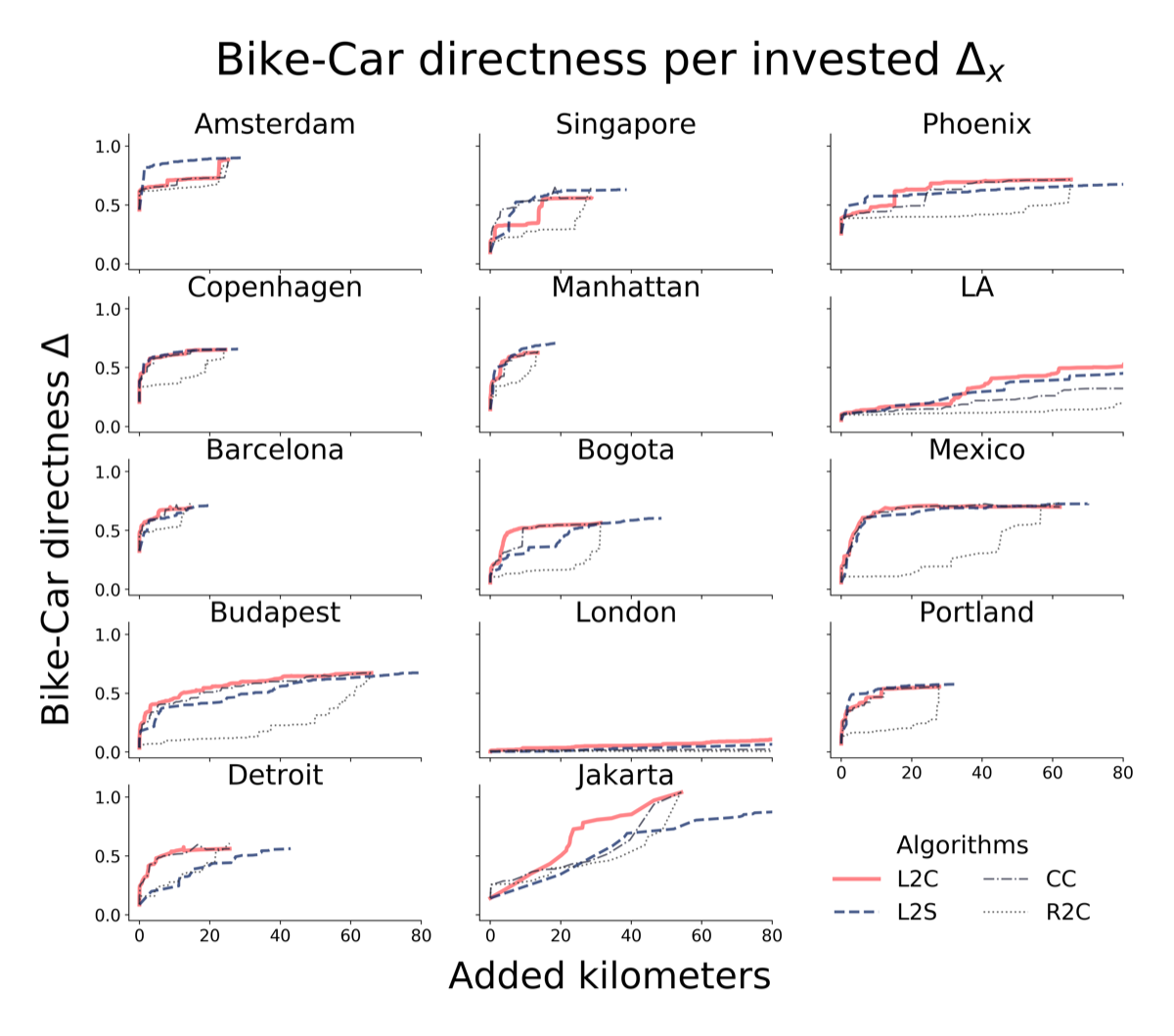}
	\caption{Bike-car directness $\Delta$ per invested kilometers.}
	\label{fig:Directness}
\end{figure*}

\begin{figure*}[h!]
	\centering
	\includegraphics[width=0.68\textwidth]{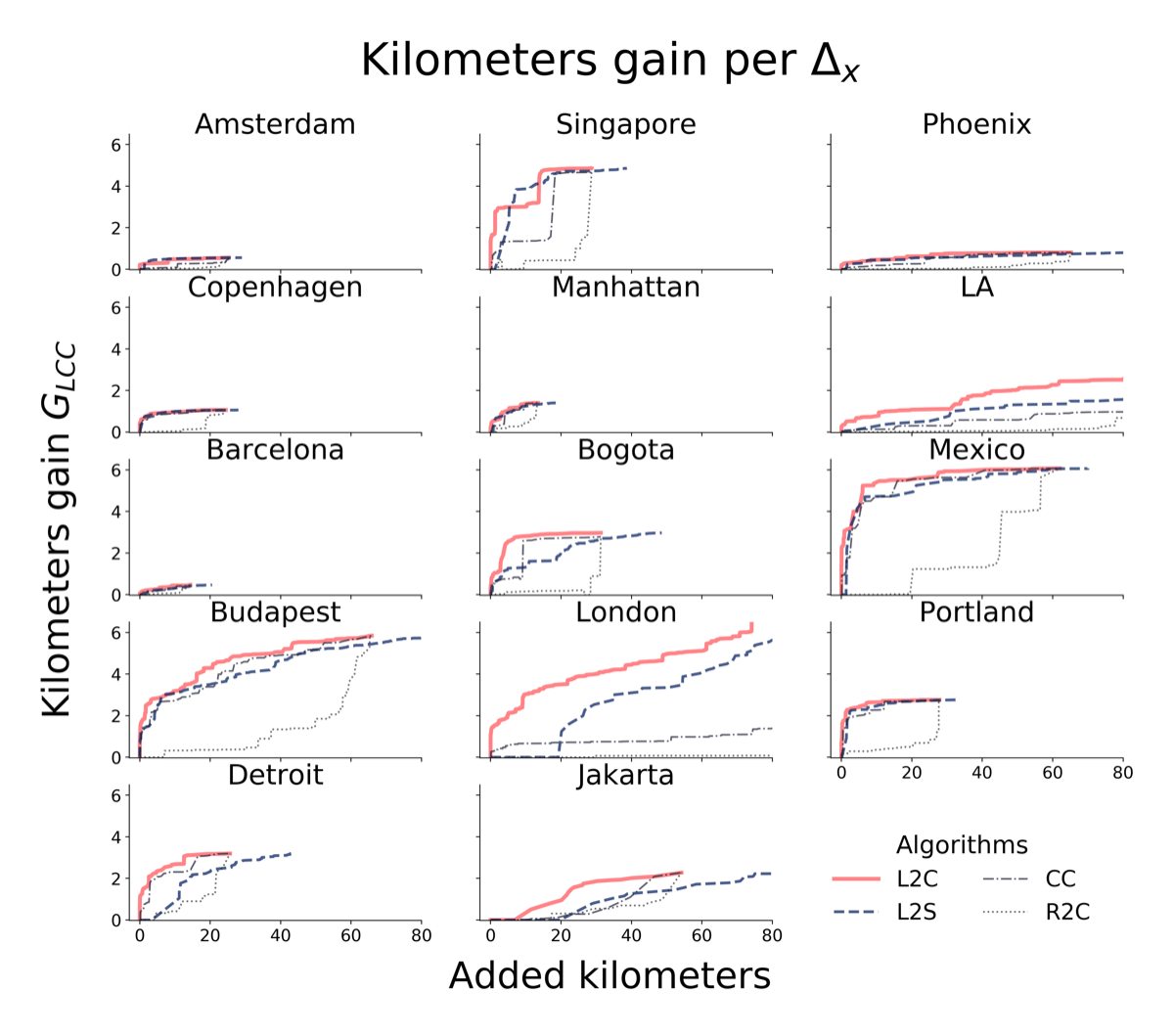}
	\caption{Kilometers gain in the largest connected component.}
	\label{fig:Gain}
\end{figure*}

Here we show the improvement of the bicycle network after the implementation of the algorithms. We measure the improvement with four different metrics. Two of them implement the notion of connectedness: i) Fraction of nodes inside the largest connected component compared to the total number of nodes in the bicycle layer, and ii) the fraction of link kilometers inside the largest connected component. In Figure \ref{fig:NodesIncrease} and \ref{fig:Lengthsncrease} we show these two measures for fourteen different cities. We also quantify iii) bicycle-to-car directness to answer the question ``how direct are the average routes of bicycles compared to cars?''. Finally, in order to measure the cumulative efficiency of our algorithms, we define the metric: iv) $G_{LCC}$ as the relative gain of bicycle path kilometers in the largest connected component. In Figures \ref{fig:Directness} and \ref{fig:Gain} we report these two measures for all algorithms and cities considered.

\end{document}